\documentclass[11pt,aps,showpacs,nofootinbib,prd,aps,epsf,floats,
               amsmath,amssymb,amsfonts]{revtex4}
\usepackage{amsmath, amssymb}
\usepackage{axodraw}
\newcommand{\mathsym}[1]{{}}

\usepackage{graphicx}
\usepackage{amsmath}
\usepackage{amssymb}
\usepackage{bm}
\newcommand{\be}{\begin{equation}}
\newcommand{\ee}{\end{equation}}
\newcommand{\bea}{\begin{eqnarray}}
\newcommand{\eea}{\end{eqnarray}}

\newcommand{\rem}[1]{}
\newsavebox{\PSLASH}
 \sbox{\PSLASH}{$p$\hspace{-1.8mm}/}
 
\renewcommand{\theequation}{\thesection.\arabic{equation}}
\newcounter{saveeqn}
\newcommand{\add}{\addtocounter{equation}{1}}
\newcommand{\alpheqn}{\setcounter{saveeqn}{\value{equation}}%
\setcounter{equation}{0}%
\renewcommand{\theequation}{\mbox{\thesection.\arabic{saveeqn}{\alph{equation}}}}}
\newcommand{\reseteqn}{\setcounter{equation}{\value{saveeqn}}%
\renewcommand{\theequation}{\thesection.\arabic{equation}}}

 \newsavebox{\notrightarrow}
 \sbox{\notrightarrow}{$\to$\hspace{-4mm}/}
 
 \newsavebox{\PARTIALSLASH}
 \sbox{\PARTIALSLASH}{$\partial$\hspace{-1.6mm}/}
 
 \newsavebox{\ASLASH}
 \sbox{\ASLASH}{$A$\hspace{-2.1mm}/}
 
 \newsavebox{\KSLASH}
 \sbox{\KSLASH}{$k$\hspace{-1.8mm}/}
 
 \newsavebox{\LSLASH}
 \sbox{\LSLASH}{$\ell$\hspace{-1.8mm}/}
 
 \newsavebox{\QSLASH}
 \sbox{\QSLASH}{$q$\hspace{-1.8mm}/}
 
 \newsavebox{\DSLASH}
 \sbox{\DSLASH}{$D$\hspace{-2.2mm}/}
 
 \newsavebox{\DbfSLASH}
 \sbox{\DbfSLASH}{${\mathbf D}$\hspace{-2.8mm}/}
 
 \newsavebox{\DELVECRIGHT}
 \sbox{\DELVECRIGHT}{$\stackrel{\rightarrow}{\partial}$}
 
 \newcommand{\blue}{\IfColor{\textCadetBlue}{}}
\newcommand{\black}{\IfColor{\textBlack}{}}
\newcommand{\red}{\IfColor{\textRed}{}}
\newcommand{\green}{\IfColor{\textOliveGreen}{}}
\newcommand{\lila}{\IfColor{\textRedViolet}{}}

\begin{document}
\title{Gauge Invariant Cutoff QED}
\author{F. Ardalan$^{a,b}$}\email{ardalan@ipm.ir}
\author{H. Arfaei$^{a,c}$}\email{arfaei@ipm.ir}
\author{M. Ghasemkhani$^{c}$}\email{ghasemkhani@ipm.ir}
\author{N. Sadooghi$^{a}$}\email{sadooghi@physics.sharif.ir}
\affiliation{$^a$Department of Physics, Sharif University of
Technology, P.O. Box 11155-9161, Tehran-Iran\\
$^{b}$Institute for Research in Fundamental Sciences (IPM)\\
School of Physics, P.O. Box 19395-5531, Tehran-Iran\\
$^{c}$Institute for Research in Fundamental Sciences (IPM)\\
School of Particles and Accelerators, P.O. Box 19395-5531,
Tehran-Iran}
\begin{abstract}
\vspace{-0.2cm}\noindent A hidden generalized gauge symmetry of a
cutoff QED is used to show the renormalizability of QED. In
particular, it is shown that corresponding Ward identities are valid
all along the renormalization group flow. The exact Renormalization
Group flow equation corresponding to the effective action of a
cutoff $\lambda\varphi^{4}$ theory is also derived. Generalization
to any gauge group is indicated.
\end{abstract}
  \maketitle
\section{Introduction}\label{introduction}
\par\noindent
Diagrammatic proof of the renormalizability of QED in the BPHZ
context is a lengthy derivation \cite{BPHZ}. In 1984, Polchinski
presented a simple proof using a version of Wilsonian
renormalization group equation (RGE) for the $\lambda \varphi^4$
theory \cite{polchinski1984}. Although Polchinski's proof of
renormalizability could in principle be extended to QED, the
manifest violation of gauge invariance in his momentum cutoff
formulation was an obstacle to a straightforward proof. This has
been extensively studied and is by now a well understood derivation
of renormalizability of QED \cite{RGapplication, gaugeinvariantRG}
(for recent reviews see \cite{pawlowski} and the literature
therein). Yet, it remains an untidy procedure involving either
complicated cutoff insertions or additional labor for the proof of
Ward identities.
\par
In this work, we present a simple extension of Polchinski's proof of
renormalizability to QED using a hidden generalized gauge invariance
of the cutoff formulation, thus circumventing the need for explicit
verification of the Ward identities. In Sec. II, we introduce our
slightly modified cutoff procedure of Polchinski and apply it to the
$\lambda \varphi^4$ theory. The cutoff procedure we use is simply
multiplying the fields by an appropriate cutoff function in the
momentum representation, as explained in Sec. II. The new point-wise
product in the momentum space, defines a ``deformed'' nonlocal
product in the coordinate space. For the reasons which will be
explained below, and are originally indicated in \cite{lizzi2012},
the cutoff function is to be taken as a sequence of analytic
functions which converge to a sharp cutoff. This procedure has its
origin in Kogut and Wilson's \cite{kogut1974} ``Incomplete
Integration'', with which Wilson's exact RGE was first formulated.
Wetterich introduced a similar concept of ``average field'' in the
discussion of RGE \cite{wetterich1991}, which was a precursor to his
well-known presentation of exact average RGE \cite{wetterich1993,
morris1993, berges2000, pawlowski}. We, however, arrived at this
cutoff procedure from an entirely different angle of translationally
invariant noncommutative gauge theory \cite{lizzi2009, ardalan2010},
which led us to the symmetry of the cutoff effective action for
gauge theories, explained in Sec. III. Here, using the new deformed
product of functions, we present the generalized (deformed) cutoff
gauge symmetry of the cutoff QED, and indicate its
renormalizability. Based on the idea proposed in the present paper,
Lizzi and Vitale show recently in \cite{lizzi2012} that the new
``deformed'' gauge symmetry, defined by a ``deformed'' product of
fields, leads to a new cocommutative Hopf algebra with ``deformed''
costructures. They argue that in order to preserve the associativity
of the new deformed product of functions, the above mentioned cutoff
function is to be analytic. Being analytic, however, the new product
can be interpreted as a simple redefinition of fields, which is
isomorphic to a point-wise product and therefore physically trivial.
To circumvent this problem, they propose the cutoff function to be a
sequence of analytic functions, which converge to a sharp cutoff, as
indicated above. Moreover, using a rigorous mathematical
construction of the new deformed Hopf algebra, they explicitly show
that the map between this deformed Hopf algebra and the standard
(undeformed) one, which is to be compatible with the suggested field
redefinition, does not satisfy the coalgebra and the Hopf algebra
homomorphisms. Thus, the new deformed gauge symmetry is inequivalent
to the standard one and defines a bona fide new symmetry. In order
to show that the proposed cutoff QED is renormalizable without
destroying the (deformed) cutoff gauge invariance, the
Ward-Takahashi's identities of the cutoff theory are to be verified.
This will be done in Sect. III. A derivation of the deformed
Ward-Takahashi's identities will be presented in Appendix A, for the
sake of completeness. As it turns out, the new deformed symmetry is
preserved along the renormalization group flow, for every fixed
ultraviolet (UV) cutoff. This was indeed expected, because as it is
pointed out in \cite{lizzi2012}, although the sharp cutoff cannot
define a deformed associative product, but the theory with a sharp
cutoff, being a limit of a Hopf gauge invariant theory, exhibits the
same symmetries of the theory defined with the deformed product of
fields.
\par
Let us notice that the cutoff procedure we use and the consequent
symmetry can be easily generalized to the gauge theory with any
group and matter field, including $SU(N)$, as the noncommutative
structure from which it is derived in Sec. III can be extended
trivially to these cases. The point being that the ``noncommutative
geometric'' structure we will be using, is in fact Abelian. For
simplicity of presentation, we will restrict ourselves to the case
of $U(1)$ gauge group, QED. In \cite{lizzi2012}, the case of $SU(N)$
gauge group is studied.
\par
In Sec. IV, we will use the idea of field deformation and the method
developed in \cite{wetterich1993, morris1993} to derive the
corresponding exact RG flow equation for the effective average
action of $\lambda\varphi^{4}$ theory including an appropriate
infrared (IR) cutoff function (see also Appendix B for a detailed
proof). We will show that the deformed flow equation is different
from the usual RG exact flow equation derived in
\cite{wetterich1993, morris1993}. The main reason for this
difference is the multiplicative nature of our cutoff function, in
contrast to the additive IR cutoff introduced in
\cite{wetterich1993,morris1993}. Modifying each field with such a
multiplicative cutoff, the cutoff function appears not only in the
kinetic term, but also in the interacting part of the classical cut
off action. It is interesting to generalize this procedure to gauge
theories and to explore the possible practical consequences of the
new exact average RG for the effective average action as well as the
hidden generalized gauge symmetry along the flow, pointed out in
this paper. The origin of this gauge symmetry will nevertheless be
discussed in the conclusions, Sec. V.
\section{Modified cutoff regularization}
\setcounter{equation}{0} \par\noindent In this section we redo
Polchinski's proof  of $\lambda \varphi^4$ theory
\cite{polchinski1984} by a slightly different cutoff procedure. The
real scalar theory defined by the action
\begin{equation}\label{G1}
S=\int d^4x~[-{\frac{1}{2}}\partial_\mu\varphi\partial^\mu\varphi +
{\frac{1}{2}}m^2\varphi^2+{\frac{1}{4!}}\lambda\varphi ^4],
\end{equation}
and the momentum cutoff $\Lambda_0$ in the Euclidean space is
regularized by Polchinski via introduction of a momentum cutoff
$K_{\Lambda_{0}}(p)$ into the propagator,
\begin{equation}\label{G2}
\frac{K_{\Lambda_{0}}(p)}{p^{2}+m^{2}}.
\end{equation}
The  main property of the function $K_{\Lambda_{0}}(p)$ is that it
has a value equal to $1$ for $p^2 < \Lambda^2 _0$ and decreases
rapidly for $\frac{p^2}{\Lambda_0^2}\rightarrow \infty$. An example
is
\begin{eqnarray}\label{G3}
K_{\Lambda_{0}}(p)&=&\left\{
\begin{array}{rclcrcl}
1,&\qquad&p^{2}\leq\frac{\Lambda_{0}^{2}}{4},\\
\exp[(1-\frac{p^{2}}{\Lambda_0^2})^{-1}\exp(4-\frac{p^{2}}{\Lambda_0^2})^{-1}],&\qquad&\frac{\Lambda_0^2}{4}<p^{2}<\Lambda_0^2,\\
0,&\qquad&p^{2}\geq \Lambda_0^2.
\end{array}
\right.
\end{eqnarray}
The cutoff function $K_{\Lambda_{0}}(p)$ effectively cuts off the
momentum integral in all loops, rendering them ultraviolet finite in
perturbation theory. The introduction of a cutoff function in the
propagators, and consequently in all loops is an efficient procedure
of implementing Wilson's renormalization group flow, as the cutoff
momentum $p$ can now be lowered all the way down to zero in the path
integral for the effective action. The remarkable aspect of the
introduction of the cutoff function in the loop integrals is that
loop integrals can now be estimated easily and cutoff independence
of Green's functions be demonstrated in a few steps in marked
contrast to the lengthy BPHZ proof of renormalizability.
\par
In this work, we propose to modify this procedure by extending the
cutoff to all the terms in the action, and in fact to all fields. In
our formulation we replace the field $\tilde{\varphi}(p)$ by their
cutoff counterpart
\begin{equation}\label{G4}
h_{\Lambda}(p)\tilde{\varphi}(p),
\end{equation}
in momentum space. Comparing to (\ref{G1}), $K_{\Lambda}$ is to be
identified with $h_{\Lambda}^{-2}$. This is therefore a
straightforward implementation of Wilson's cutoff procedure executed
directly on the fields rather than on the path integrals.
\par
In the original Polchinski's formulation, cutoff independence of the
theory is
\begin{equation}\label{G6}
\frac{\partial {\cal Z}}{\partial t}=0, \qquad\mbox{with}\qquad
\frac{\partial}{\partial t}\equiv
\Lambda\frac{\partial}{\partial\Lambda},
\end{equation}
and, the running effective action is
\begin{equation}\label{G7}
S_{\mbox{\tiny{eff}}}=\int\frac{d^{4}p}{(2\pi)^{4}}\bigg[-\frac{1}{2}\tilde{\varphi}(p)\frac{p^{2}+m^{2}}{K_{\Lambda}(p)}\tilde{\varphi}(-p)+{\cal
L}_{\mbox{\tiny{int}}}(\tilde{\varphi},\Lambda)\bigg].
\end{equation}
Equation (\ref{G6}) then determines the running effective
interaction lagrangian ${\cal L}_{\mbox{\tiny{int}}}$ at the
$\Lambda$ scale, which now satisfies the functional differential
equation
\begin{equation}\label{G8}
\frac{\partial{\cal L}_{\mbox{\tiny{int}}}}{\partial
t}=-\frac{1}{2}\int\frac{d^{4}p}{(2\pi)^{4}}\frac{1}{p^{2}+m^{2}}~\frac{\partial
K_{\Lambda}(p)}{\partial t}\bigg[\frac{\partial {\cal
L}_{\mbox{\tiny{int}}}}{\partial\tilde{\varphi}(p)}\frac{\partial{\cal
L}_{\mbox{\tiny{int}}}}{\partial\tilde{\varphi}(-p)}+\frac{\partial^{2}{\cal
L}_{\mbox{\tiny{int}}}}{\partial\tilde{\varphi}(p)\partial\tilde{\varphi}(-p)}\bigg].
\end{equation}
Our formulation does not yet differ from Polchinski's as the
functional ${\cal L}_{\mbox{\tiny{int}}}$ is not yet specified. The
difference shows up when expanding ${\cal L}_{\mbox{\tiny{int}}}$ in
a series in $\tilde{\varphi}$.
\begin{equation}\label{G9}
{\cal
L}_{\mbox{\tiny{int}}}(\tilde{\varphi},\Lambda)=\sum_{n=1}^{\infty}\frac{1}{(2n)!}\int\frac{d^{4}p_{1}\cdots
d^{4}p_{2n}}{(2\pi)^{8n-4}}~
L_{2n}(p_{1},\cdots,p_{2n};\Lambda)\delta(\sum_{i}p_{i})~\tilde{\varphi}(p_{1})\cdots\tilde{\varphi}
(p_{2n}).
\end{equation}
In Polchinski's formulation the renormalization flow equation
(\ref{G8}) reduces to a set of equations for the coefficient
functions $L_{2n} (p_{1},\cdots,p_{2n},\Lambda)$, whose solution
would give the effective action at scale $\Lambda$,
\begin{eqnarray}\label{G10}
\lefteqn{(\frac{\partial}{\partial t}+4-2n)
L_{2n}(p_{1},\cdots,p_{2n};\Lambda)}\nonumber\\
&=&-\sum_{\ell=1}^{n}\bigg\{ Q_{\Lambda}(P,m)
L_{2\ell}(p_{1},\cdots,p_{2\ell-1};\Lambda)L_{2n+2-2\ell}(p_{2\ell},\cdots,p_{2n},-P;\Lambda)+permutation\nonumber\\
&&\hspace{1.2cm}-\frac{1}{2}
\int\frac{d^{4}p}{(2\pi\Lambda)^{4}}L_{2n+2}(p_{1},\cdots,p_{2n},P,-P;\Lambda)
Q_{\Lambda}(P,m)\bigg\}.
\end{eqnarray}
Here, $P=\sum_{i=1}^{2\ell-1}p_{i}$, and
\begin{equation}\label{G11}
Q_{\Lambda}(P,m)=\frac{1}{P^{2}+m^{2}}~\Lambda^{2}\frac{\partial
K_{\Lambda}(P)}{\partial t}.
\end{equation}
A convenient rescaling $L_{2n}\rightarrow \Lambda^{4-2n} L_{2n}$ has
been inserted in (\ref{G10}).
\par
In our formulation, however, the expansion in (\ref{G9}) has to be
replaced by
\begin{equation}\label{G12}
{\cal
L}_{\mbox{\tiny{int}}}(\tilde{\varphi},\Lambda)=\sum_{n=1}^{\infty}\frac{1}{(2n)!}\int\frac{d^{4}p_{1}\cdots
d^{4}p_{2n}}{(2\pi)^{8n-4}}~
L'_{2n}(p_{1},\cdots,p_{2n};\Lambda)\delta(\sum_{i}p_{i})~h_{\Lambda}(p_{1})\tilde{\varphi}(p_{1})\cdots
h_{\Lambda}(p_{2n})\tilde{\varphi}(p_{2n}),
\end{equation}
as every momentum space field $\tilde{\varphi}(p)$ is rescaled and
cut off by the cutoff function $h_{\Lambda} (p)$ from (\ref{G4}). It
may seem that the subsequent steps in renormalizality may get
complicated, but this is not the case. In fact the resulting
renormalization group flow equations for the function
$h_{\Lambda}(p_{1})\cdots
h_{\Lambda}(p_{2n})L'_{2n}(p_{1},\cdots,p_{2n};\Lambda)$, now would
be exactly the same form as (\ref{G10}) for the coefficient
functions, where the function $K_{\Lambda} (p)$ of Polchinski has to
be replaced by $h_{\Lambda}^{-2} (p)$, see \cite{polchinski1984} for
more details. The advantage of Polchinski's approach to
renormalization is that the insertion of a cutoff function in the
integrals allows him to estimate the coefficient functions
$L_{2n}(p_{1},\cdots,p_{2n})$ through functional analytic methods
and put bounds on them; and then to prove cutoff independence of the
Green's functions in perturbation theory, through a series of
Lemmas. Note that in all this the function $K_{\Lambda}$ appears, as
in (\ref{G8}), only in its derivative form
$\partial_{t}K_{\Lambda}$, which does not affect the estimate
arguments which involve various norms of functions. We will not go
through the entire analysis, but only point out the validity of the
procedure in our approach in cases where there is ground for doubt.
\par
The first instance that our approach may invalidate Polchinski's
result is when a bound on $L'_{2n}(p_{1},\cdots,p_{2n})$ is obtained
through (\ref{G10}) using the bounds
\begin{equation}\label{G13}
\int\frac{d^{4}p}{(2\pi)^{4}}|Q_{\Lambda}(p,m)|<C\Lambda^{4},
\end{equation}
and
\begin{equation}\label{G14}
\mbox{max}~\bigg|\frac{\partial^{n}}{\partial
p^{n}}Q_{\Lambda}(p,m)\bigg|<D_{n}\Lambda^{-n}.
\end{equation}
Here, $C$ and $D_{n}$ are appropriate constants. The result
\begin{eqnarray}\label{G15}
\begin{array}{rclcccc}
\mbox{max}~\bigg|L_{2n}^{(r)}(p_{1},...,p_{2n};\Lambda)\bigg|&\leq&{\cal
P}_{2r-n}(\ln\frac{\Lambda_{0}}{\Lambda}),&\qquad&\mbox{for}&\qquad&r+1-n\geq
0,\\
&=&0,&\qquad&\mbox{for}&\qquad&r+1-n<0,
\end{array}
\end{eqnarray}
and similar bounds are central to the proof of renormalizability. In
(\ref{G15}), $L_{2n}^{(r)}$ is the r'th term in the perturbative
expansion of $L_{2n}$, and ${\cal P}_{2r-n}$ are polynomials of
order $2r-n$.
\par
In our case, (\ref{G10}) involves $h_{\Lambda} (p_1)\cdots
h_{\Lambda}(p_n)L_{2n}$ rather than $L_{2n}$. One must make sure
that the appearance of the $h_{\Lambda}$'s does not ruin the bounds
in Polchinski and therefore ruin the arguments on renormalizability
of the theory. We have identified the function $K_{\Lambda}$ of
(\ref{G2}) with $h_{\Lambda}^{-2}$ of (\ref{G4}), as mentioned
above. Therefore if $K_{\Lambda}$'s should behave as in Polchinski's
formulation, i.e. go to zero when $p^{2}$ approaches $\Lambda^{2}$,
and vanish for $p^{2}>\Lambda^{2}$, then $h_{\Lambda}$ should become
large as $p^{2}$ approaches $\Lambda^{2}$. Of course, we set
$h_{\Lambda}(p^{2})=0$ for $p^{2}>\Lambda^{2}$. Therefore in the
left hand side (l.h.s.) of (\ref{G10}), after inserting
$h_{\Lambda}\cdots h_{\Lambda}L'_{2n}$ for $L_{2n}$'s, as the
$h_{\Lambda}$'s are larger than one in their range of definition,
they can be dropped in the ensuing inequality. On the right hand
side (r.h.s.) of (\ref{G10}), we now use the functional norm with an
appropriate weight to kill off the value of $h_{\Lambda}>1$ for
$p^{2}\to\Lambda^{2}$,\footnote{See footnote on page 280 of
Polchinski \cite{polchinski1984}.} leading to the desired inequality
(\ref{G15}), this time for $L'_{2n}$'s. Let us note that the above
choice for $K_{\Lambda}$, is not a unique one. It is easy to show,
that Polchinski's proof is also correct for $K_{\Lambda}$ becoming
large as $p^{2}$ approaches $\Lambda^{2}$. In the next section, we
will use this second alternative, and define, as in
\cite{lizzi2012}, the cutoff function $h_{\Lambda}(p)$ as a sequence
of analytic functions satisfying
\begin{eqnarray}\label{G16}
h_{\epsilon,\Lambda}(p)\stackrel{\epsilon\to
0}{\longrightarrow}\Theta_{\Lambda}(p)=\left\{
\begin{array}{ccccrcl}
0&\qquad&\mbox{for}&\qquad&p^{2}&\geq& \Lambda^{2},\\
1&\qquad&\mbox{for}&\qquad&p^{2}&<&\Lambda^{2}.
\end{array}
\right.
\end{eqnarray}
Here, $\Theta_{\Lambda}(p)$ is a sharp (UV) cutoff.\footnote{The
subscript $\epsilon$ on $h_{\epsilon,\Lambda}(p)$ will be omitted in
the rest of this paper.} This property is necessary to avoid the
interpretation of the deformation (\ref{G4}) to be just a
redefinition of fields in the momentum space \cite{lizzi2012} (for
more mathematical details, see the discussions at the end of the
next section).
\section{Gauge invariance of cutoff QED and Ward identities}
\setcounter{equation}{0}
\par\noindent
The Polchinski's procedure was applied to QED very early and
provided a simple proof of renormalizability of the theory
\cite{RGapplication}. In QED there are now two propagators to cut
off, that of the electron and that of the photon. This can be done
with the same cutoff function. The renormalization group equation is
similarly derived and estimates for the coefficient functions and
bounds on Green's functions obtained. The situation is then a
straightforward extension of the $\lambda \varphi^4$ theory. There
is only one significant hurdle to overcome which has engaged authors
of Refs. \cite{RGapplication, gaugeinvariantRG} ever since and is
the subject of the present work, \textit{i.e.} the question of gauge
invariance.
\par
The problem is that Polchinski's approach, and in fact any approach
involving a momentum cutoff, inherently violates gauge invariance:
Gauge invariance is a statement about the behavior of gauge fields
in a space-time point, involving all momenta. Thus, at any finite
cutoff scale $\Lambda$, the flow equation and its solutions are not
gauge invariant. However, it was proved, that the final IR point of
the flow $\Lambda \rightarrow 0 $, the expressions for the quantum
effective action and the Green's functions are indeed invariant
\cite{gaugeinvariantRG}. There were also nontrivial modifications of
the cutoff procedure which were not gauge invariant all along the
flow, but only at its end points. These formulations have been
extensively pursued in the application of the exact renormalization
group in such areas as QCD (see \cite{pawlowski} for recent
reviews), and gravity \cite{RGgravity}. In all these works
derivation of the modified Ward-Takahashi identities is the
essential complication.
\par
In this section, we will show that our version of introduction of
the momentum cutoff in the theory ensures persistence of gauge
invariance in the form of a generalized deformed symmetry of the
cutoff QED and derive the resultant deformed Ward-Takahashi
identities in a standard manner. We start from the classical action
of QED,\footnote{In QED, the ghost fields decouple from the theory.}
\begin{equation}\label{H1}
S_{\tiny\mbox{QED}}=\int
d^{4}x\bigg\{\bar\psi(i\gamma^{\mu}\partial_{\mu}-m)\psi-e\bar\psi\gamma^{\mu}A_{\mu}\psi-
\frac{1}{4}F_{\mu\nu}F^{\mu\nu}-\frac{1}{2\xi}(\partial_{\mu}A^{\mu})^{2}\bigg\},
\end{equation}
with the cutoff function $h_{\Lambda} (p)$ inserted on each field in
the momentum space
\begin{eqnarray}\label{H1-b}
\tilde{\psi}(p)\rightarrow h_{\Lambda}(p)\tilde{\psi}(p),\qquad
\mbox{and}\qquad \tilde{A}_{\mu}(p)\rightarrow
h_{\Lambda}(p)\tilde{A}_{\mu}(p).
\end{eqnarray}
We arrive at the deformed cutoff (effective) action
\begin{eqnarray}\label{H1-c}
S_{h_\Lambda}&=&-\frac{1}{2}\int\frac{d^{4}k}{(2\pi)^{4}}~h^{2}_{\Lambda}(k)
\bigg[k^{2}g_{\mu\nu}-\left(1-\frac{1}{\xi}\right)k_{\mu}k_{\nu}\bigg]\tilde{A}^{\mu}(k)\tilde{A}^{\nu}(-k)\nonumber\\
&&+\int
\frac{d^{4}p}{(2\pi)^{4}}~h_{\Lambda}^{2}(p)\tilde{\bar{\psi}}(p)\left(\gamma_{\mu}p^{\mu}-m\right)\tilde{\psi}(-p)\nonumber\\
&&-e\int
\frac{d^{4}p}{(2\pi)^{4}}\frac{d^{4}q}{(2\pi)^{4}}\frac{d^{4}\ell}{(2\pi)^{4}}~h_{\Lambda}(p)h_{\Lambda}(\ell)h_{\Lambda}(q)\tilde{\bar{\psi}}(p)\gamma^{\mu}
\tilde{A}_{\mu}(\ell)\tilde{\psi}(q)\delta^{4}(\ell+q-p).
\end{eqnarray}
Here, the cutoff function $h_{\Lambda} (p)$ is to be analytic and
has to converge to a sharp (UV) cutoff function, as is expressed in
\eqref{G16}. Moreover, it has to satisfy
$h_{\Lambda}(-p)=h_{\Lambda}(p)$. The reason for this specific
choice will be explained in what follows. But before doing this, let
us consider the effective action (\ref{H1-c}). As it turns out, it
has a symmetry which is the generalization of gauge symmetry of
$S_{\tiny\mbox{QED}}$. Whereas the gauge symmetry of
$S_{\tiny\mbox{QED}}$ is
\begin{equation}\label{H2}
\psi(x)\rightarrow e^{ie\epsilon(x)}\psi(x),\qquad
\bar\psi(x)\rightarrow e^{-ie\epsilon(x)}\bar\psi(x),\qquad
A_{\mu}\rightarrow A_{\mu}(x)-\partial_{\mu}\epsilon(x),
\end{equation}
the symmetry of $S_{h_{\Lambda}}$ is similarly defined but now
involves $h_{\Lambda}$. To introduce this new (deformed) cutoff
gauge symmetry of $S_{h_{\Lambda}}$, let us first notice that when
in momentum space two functions $\tilde{\psi}(p)$ and
$h_{\Lambda}(p)$ are point-wise multiplied, their corresponding
functions in the configuration space, $\psi(x)$ and
$\tilde{h}_{\Lambda}(x)$, are multiplied via convolution,
\begin{equation}\label{ H3}
h_{\Lambda}(p)\tilde{\psi}(p)\rightarrow\tilde{h}_{\Lambda}(x)\circ\psi(x),
\end{equation}
where convolution of two functions $f(x)$ and $g(x)$ is defined by,
\begin{equation}\label{H4}
(f\circ g)(x)\equiv\int d^{4}y~ f(x-y)g(y).
\end{equation}
The above mentioned deformed gauge symmetry transformation of
$S_{h_{\Lambda}}$ is then given by
\begin{equation}\label{H5}
\psi(x)\rightarrow(\widetilde{h^{-1}_{\Lambda}})\circ[(\tilde{h}_{\Lambda}\circ
g)(\tilde{h}_{\Lambda}\circ\psi)].
\end{equation}
Here, $g(x)$ is the generalization of $e^{ie\epsilon(x)}$ defined by
\begin{equation}\label{H6}
g(x)=1+ie\epsilon(x)+\frac{1}{2!}(\widetilde{h^{-1}_{\Lambda}})\circ[(\tilde{h}_{\Lambda}\circ
(ie\epsilon))(\tilde{h}_{\Lambda}\circ (ie\epsilon))]+\cdots
\end{equation}
The transformation of $A_{\mu}(x)$ is
\begin{equation}\label{H7}
A_{\mu}(x)\rightarrow
A_{\mu}(x)+(\widetilde{h_{\Lambda}^{-1}})\circ\bigg[\left(\tilde{h}_{\Lambda}\circ
g\right)\left(\tilde{h}_{\Lambda}\circ
(\partial_{\mu}g^{-1})\right)\bigg].
\end{equation}
These strange looking transformations come from a simple generalized
noncommutative geometric construction, related to the
translationally invariant noncommutative star-product, introduced
originally in \cite{lizzi2009}. to understand the origin of the
above deformed gauge transformations (\ref{H5})-(\ref{H7}), we
review, in what follows, this generalized noncommutative field
theory.
\par
Let us start by defining the generalized translationally invariant
noncommutative star-product from \cite{lizzi2009, ardalan2010}, as a
generalization of the usual $C^{\star}$-algebra of point-wise
multiplication algebra of functions,
\begin{equation}\label{A1}
(f\star g)(x)\equiv
\int\frac{d^{4}p}{(2\pi)^{4}}\frac{d^{4}q}{(2\pi)^{4}}~e^{ipx}{\cal
K}(p,q)\tilde{f}(p-q)\tilde{g}(q).
\end{equation}
The point-wise multiplication is the special case of ${\cal K}=1$.
Associativity of the algebra is the main constraint on the function
${\cal K}$. It is
\begin{equation}\label{A2}
{\cal K}(p,q){\cal K}(q,r)={\cal K}(p,r){\cal K}(p-r,q-r).
\end{equation}
It was shown in \cite{ardalan2010} that the following expression is
a solution of (\ref{A2})
\begin{equation}\label{A3}
{\cal{K}}(p,q)=h^{-1}(p)h(q)h(p-q)e^{i\Omega(p,q)},
\end{equation}
with
\begin{equation}\label{A4}
\Omega(p,q)=\theta_{\mu\nu}p^{\mu}q^{\nu}+\eta(q)-\eta(p)+\eta(p-q).
\end{equation}
Here, $\theta_{\mu \nu}$ is an antisymmetric constant matrix, $h(p)$
and $\eta(p)$ are arbitrary real even and odd functions,
respectively. Later, $h(p)$, will be identified with the cutoff
function $h_{\Lambda}(p)$ satisfying the properties \eqref{G16} and
converging to a sharp cutoff function $\Theta_{\Lambda}(p)$. It is
readily seen that for $\theta \neq 0$, the algebra is
noncommutative, and it is commutative when $\theta=0$. When
$\theta=0$ and $\eta=0$, the new star-product in the momentum space,
involves multiplications of the functions of the algebra by the
fixed function $h(p)$
\begin{equation}\label{A5}
(f\star
g)(x)=\int\frac{d^{4}p}{(2\pi)^{4}}~\frac{1}{h(p)}e^{ipx}\int\frac{d^{4}q}{(2\pi)^{4}}[h(p-q)\tilde{f}(p-q)][h(q)\tilde{g}(q)].
\end{equation}
Thus
\begin{eqnarray}\label{A6}
h(p)(\widetilde{f\star
g})(p)&=&\int\frac{d^{4}q}{(2\pi)^{4}}[h(p-q)\tilde{f}(p-q)][h(q)\tilde{g}(q)]\nonumber\\
&=&[(h\tilde{f})\circ(h\tilde{g})](p),
\end{eqnarray}
where the convolution of two functions $f$ and $g$ in coordinate
space is defined in (\ref{H4}). Using (\ref{A6}), the product
(\ref{A5}) is defined as
\begin{eqnarray}\label{A7}
(f\star g)(x)=(\widetilde{h^{-1}})\circ[(\tilde{h}\circ
f)(\tilde{h}\circ g)](x).
\end{eqnarray}
Here, we have also used the relation
\begin{equation}\label{A8}
(\widetilde{fg})=({\tilde{f}\circ \tilde{g}}).
\end{equation}
It must be mentioned that as long as $h$ is a smooth one-to-one map,
the new star-algebra is an isomorphism of the algebra with the
$C^{\star}$-algebra. However, there is nothing to forbid the
function $h$ to be singular. In fact, we have used such a singular
$h$ to cut off the momentum of the field theory. From now on, we
will identify $h$ with $h_{\Lambda}$, satisfying \eqref{G16} and
converging to a sharp cutoff function $\Theta_{\Lambda}(p)$
\cite{lizzi2012}. In our case the function $h_{\Lambda}$ effectively
cuts down the domain of the function space. The inverse function
$h_{\Lambda}^{-1}$ appearing above should be understood in this
context. Indeed, we do not need to worry about the appearance of
$h_{\Lambda}^{-1}$ in the definition \eqref{A1}. Using
$h_{\Lambda}(p=0)=1$ in \eqref{A5}, with $h$'s identified with
$h_{\Lambda}$'s, we have
\begin{eqnarray}\label{A13}
\int d^{4}x (f\star g)(x)&=&\int
\frac{d^{4}q}{(2\pi)^{4}}~[h_{\Lambda}(-q)\tilde{f}(-q)][h_{\Lambda}(q)\tilde{g}(q)]\nonumber\\
&=&\int
\frac{d^{4}q}{(2\pi)^{4}}\frac{d^{4}p}{(2\pi)^{4}}\delta^{4}(p+q)[h_{\Lambda}(p)\tilde{f}(p)][h_{\Lambda}(q)\tilde{g}(q)].
\end{eqnarray}
Hence, as it turns out, the integrated form of the star-product of
two functions, $f$ and $g$, in the coordinate space can be
understood as modifying the Fourier transformed of these functions
with a ``cutoff function'' $h_{\Lambda}$, i.e.,
\begin{eqnarray*}
\tilde{f}(p)\to h_{\Lambda}(p)\tilde{f}(p),\qquad\mbox{and}\qquad
\tilde{g}(p)\to h_{\Lambda}(p)\tilde{g}(p).
\end{eqnarray*}
Using \eqref{A13}, it is easy to show that the effective action of
cutoff QED from \eqref{H1-c}, can be given in terms of the
translationally invariant star-product \eqref{A1} with the specific
choice of $\theta=0=\eta$ in \eqref{A4},
\begin{eqnarray}\label{A9}
S_{h_{\Lambda}}&=&\int
d^{4}x\bigg\{\bar\psi(i\gamma^{\mu}\partial_{\mu}-m)\star\psi-e\bar\psi\star\gamma^{\mu}A_{\mu}\star\psi-
\frac{1}{4}F_{\mu\nu}\star
F^{\mu\nu}-\frac{1}{2\xi}(\partial_{\mu}A^{\mu})\star(\partial_{\nu}A^{\nu})\bigg\},\nonumber\\
\end{eqnarray}
where $F_{\mu\nu}\equiv\partial_{\mu}A_{\nu}-\partial_{\nu}A_{\mu}$.
This is a specific version of translationally invariant
``noncommutative'' QED introduced in \cite{ardalan2010}.\footnote{In
this work, we have limited ourselves to the special case, albeit
commutative, of $\theta=0$, $\eta = 0$ in (\ref{A4}). With this
special choice of $\theta$ and $\eta$, the field strength tensor
$F_{\mu\nu}$ turns out to be the same as in commutative QED. The
ghosts will also decouple from the theory.} The above action
$S_{h_{\Lambda}}$ is invariant under the star-gauge transformation
\begin{equation}\label{A11}
\psi\rightarrow g\star\psi,\qquad\mbox{and}\qquad A_{\mu}\rightarrow
g\star A_{\mu}\star g^{-1}-\frac{i}{e}g\star\partial_{\mu}g^{-1},
\end{equation}
where
\begin{equation}\label{A12}
g=e^{ie\epsilon}_{\star}=1+ie\epsilon+\frac{1}{2!}(ie\epsilon)\star(ie\epsilon)+\cdots,
\end{equation}
(see \cite{lizzi2009, ardalan2010} for more details). Using
\eqref{A7}, it is easy to show that the transformation of $\psi$ in
(\ref{H5}) is the same as $\psi\rightarrow g\star\psi$ appearing in
(\ref{A11}), and the gauge transformation of $A_{\mu}$ in (\ref{H7})
is equivalent with the gauge transformation $A_{\mu}\rightarrow
g\star A_{\mu}\star g^{-1}-\frac{i}{e}g\star\partial_{\mu}g^{-1}$,
appearing in \eqref{A11}. Moreover, (\ref{H6}) can be identified
with (\ref{A12}).  We note that the relations (\ref{H5}) and
(\ref{H7}), which are not local anymore, reduce to the usual gauge
transformations of QED when $h_{\Lambda}= 1$. These generalized
(deformed) ``cutoff gauge transformations'' are therefore shown to
be the symmetry of $S_{h_{ \Lambda}}$. They ensure the ``good''
properties of the effective action $S_{h_{ \Lambda}}$ all along the
renormalization group flow, including transversality of the photon,
and exclusion of unwanted non-gauge invariant terms in the effective
action $S_{h_{ \Lambda}}$.
\par
We will now write down the consequent Ward-Takahashi identities. To
be complete, we will present their derivations in App. A. As it
turns out, there are essentially two different well-known such
identities which take the following forms in our case,
\begin{eqnarray}\label{H8}
&&\hspace{-1.4cm}-\frac{1}{\xi}p^{2}p_{\mu}\frac{\delta
W}{\delta\tilde{J}_{\mu}(-p)}+h^{2}_{\Lambda}(p)p_{\mu}\tilde{J}^{\mu}(p)
\nonumber\\
&&\hspace{-0.9cm}-eh_\Lambda(p)
\int\frac{d^{4}q}{(2\pi)^{4}}\bigg[h_{\Lambda}(q)h_{\Lambda}^{-1}(q-p)\frac{\delta
W}{\delta\tilde{\chi}(q-p)}\tilde{\chi}(q)+h_{\Lambda}^{-1}(q)h_{\Lambda}(q-p)\tilde{\bar{\chi}}(q-p)\frac{\delta
W}{\delta\tilde{\bar{\chi}}(q)} \bigg]=0,\nonumber\\
\end{eqnarray}
for the generating function $W[J_{\mu},\chi,\bar{\chi}]$ of
connected Green's function, and
\begin{eqnarray}\label{H9}
&&\hspace{-1.4cm}-\frac{1}{\xi}p^{2}p_{\mu}h^{2}_{\Lambda}(p)A^{\mu}(p)-
p_{\mu}\frac{\delta\Gamma}{\delta\tilde{A}_{\mu}(-p)}
\nonumber\\
&&\hspace{-0.9cm}+eh_\Lambda(p)
\int\frac{d^{4}q}{(2\pi)^{4}}\bigg[h^{-1}_{\Lambda}(q)h_{\Lambda}(q-p)\tilde{\bar{\psi}}(q-p)
\frac{\delta\Gamma}{\delta\tilde{\bar{\psi}}(q)}
+h_{\Lambda}(q)h^{-1}_{\Lambda}(q-p)\frac{\delta\Gamma}{\delta\tilde{\psi}(q-p)}\tilde{\psi}(q)
\bigg]=0,\nonumber\\
\end{eqnarray}
for the generating functional $\Gamma[A_{\mu},\bar{\psi},\psi]$ one
particle irreducible (1PI) vertex functions. It is known that
$W[J_{\mu},\chi,\bar{\chi}]$ and $\Gamma[A_{\mu},\bar{\psi},\psi]$
are related through the Legendre transformation
\begin{eqnarray}\label{H10}
\Gamma[\tilde{A}_{\mu},\tilde{\bar{\psi}},\tilde{\psi}]=W[\tilde{J}_{\mu},\tilde{\chi},\tilde{\bar{\chi}}]-\int
\frac{d^{4}p}{(2\pi)^{4}}h_{\Lambda}^{2}(p)\bigg[\tilde{\bar{\chi}}(p)\bar{\psi}(p)+\tilde{\bar{\psi}}(p)\tilde{\chi}(p)+\tilde{J}_{\mu}(p)\tilde{A}^{\mu}(-p)\bigg],
\end{eqnarray}
with respect to the sources
$(\tilde{J}_{\mu},\tilde{\chi},\tilde{\bar{\chi}})$ corresponding to
the fields $(\tilde{A}_{\mu},\tilde{\bar{\psi}},\tilde{\psi})$ in
momentum space. The Ward identities (\ref{H8}) and (\ref{H9})
correspond to the usual Ward identities of QED with identical
physical significance establishing the r$\hat{\mbox{o}}$le of our
generalized ``cutoff gauge invariance'' in the renormalization group
flow and the renormalization program of QED. They also guarantee the
invariance of the full quantum action under the (deformed) cutoff
gauge invariance (\ref{H5})-(\ref{H7}).
\par
At this stages some remarks on the properties of the cutoff function
$h_{\Lambda}(p)$ are in order.  First, let us note that
$h_{\Lambda}(p)$ is to be an analytic function. Otherwise, the
algebra defined by the translationally invariant star-product
(\ref{A1}) is not associative [see (\ref{A2}) for the associativity
condition of our deformed product]. On the other hand, taking
$h_{\Lambda}(p)$ as an arbitrary analytic function without requiring
that it converges to a sharp cutoff function, may lead to the
interpretation that the proposed deformation (\ref{H1-b}) of fields
is just a simple field redefinition, which is isomorphic to a
point-wise product and therefore physically trivial
\cite{lizzi2012}. To bypass this apparent discrepancy, the cutoff
function $h_{\Lambda}(p)$ is to be chosen as a sequence of
analytical functions which converge to a sharpf UV cutoff
$\Theta_{\Lambda}(p)$ [see \eqref{G16}]. This is recently suggested
by Lizzi and Vitale in \cite{lizzi2012}. Based on the ideas proposed
in the present paper, they show that the new deformed product of
fields leads to a new cocommutative Hopf algebra with deformed
costructure. Using a rigorous mathematical construction, they also
show that taking a cutoff function $h_{\Lambda}(p)$ that satisfies
(\ref{G16}) guarantees that the deformed Hopf algebra is
inequivalent with the standard (undeformed) Hopf algebra, and that
the new (deformed) cutoff gauge invariance, (\ref{H5})-(\ref{H7}),
is indeed an authentic new gauge symmetry. Note that the fact that
$h_{\Lambda}(p)$ converges to $\Theta_{\Lambda}(p)$ guarantees the
invariance of the cutoff action (\ref{H1-c}) under the new
(deformed) cutoff gauge invariance at each step of the limiting
procedure. The question whether this symmetry is destroyed by
renormalization is negated by the explicit proof of Ward-Takahashi's
identities, which seems to arise from the standard Ward identities
by a simple redefinition of fields $\grave{\mbox{a}}$ la
(\ref{H1-b}).
\section{Exact RG flow equation for the effective average action of cutoff $\lambda\varphi^{4}$
theory}\label{sec3}
\setcounter{equation}{0}
\par\noindent
In a separate development an alternative renormalization group
equation was derived for the effective average action
\cite{wetterich1993, morris1993, berges2000, pawlowski} to be
defined below. The idea was to add an IR cutoff term
\begin{eqnarray}\label{S1}
\Delta {\cal{S}}_{k}[\varphi]=\frac{1}{2}\int
\frac{d^{4}q}{(2\pi)^{4}}{\cal{R}}_{k}(q)
\tilde{\varphi}(-q)\tilde{\varphi}(q),
\end{eqnarray}
to the classical action (\ref{G1}) in the Euclidean space, and to
modify in this way the standard effective action of the theory. In
(\ref{S1}), the IR cutoff ${\cal{R}}_{k}$ satisfies the following
properties
\begin{eqnarray}\label{S2}
{\cal{R}}_{k}(q)\left\{\begin{array}{cccrcl}
=0&&\mbox{for}&k&\to&0\\
\to \infty&&\mbox{for}&k&\to& \Lambda~~\mbox{or}~~k\to\infty.
\end{array}
\right.
\end{eqnarray}
An example for ${\cal{R}}_{k}(q)$, which is also used in
\cite{berges2000} is
\begin{eqnarray}\label{S3}
{\cal{R}}_{k}(q)\sim\frac{q^{2}}{e^{\frac{q^{2}}{k^{2}}}-1},
\end{eqnarray}
that behaves as ${\cal{R}}_{k}(q)\sim k^{2}$ for fluctuations with
small momenta $q^{2}\ll k^{2}$, and vanishes for $q^{2}\gg k^{2}$.
Adding $\Delta {\cal{S}}_{k}[\varphi]$ to the classical action and
integrating over all fluctuations to derive the effective action of
the theory will induce automatically an effective mass $\sim k$ to
those Fourier modes of $\tilde{\varphi}(q)$ with small momenta
$q^{2}\ll k^{2}$, prohibiting them from contributing to the
effective average action of the theory, $\Gamma_{k}$. The resulting
effective average action $\Gamma_{k}[\phi]$ will depend on the scale
$k$ and satisfies the RG flow equation [see e.g.
\cite{wetterich1993, morris1993, berges2000, pawlowski} for a
rigorous derivation of (\ref{S4})]
\begin{eqnarray}\label{S4}
\partial_{t}\Gamma_{k}[\phi]=\frac{1}{2}\mbox{Tr}\bigg\{G_{k}^{(2)}
\partial_{t}{\cal{R}}_{k}\bigg\},
\end{eqnarray}
where $\phi\equiv \langle\varphi\rangle$ and $\partial_{t}\equiv
k\partial_{k}$. The trace involves an integration over momenta.
Moreover, $G^{(2)}_{k}$ is the full connected two-point Green's
function satisfying
\begin{eqnarray}\label{S4a}
G^{(2)}_{k}=[\Gamma_{k}^{(2)}+{\cal{R}}_{k}]^{-1}.
\end{eqnarray}
Here, $\Gamma_{k}^{(2)}$ is the exact of 1PI two-point vertex
function, arising from variation of the effective average action
$\Gamma_{k}[\phi]$ two times with respect to $\phi$. By definition,
the effective average  action interpolates between the classical
action, $\Gamma_{\Lambda}\approx S_{0}$, and the full effective
action $\Gamma=\lim\limits_{k\to 0}\Gamma_{k}$ \cite{berges2000}. In
$\Gamma_{\Lambda}$, $\Lambda$ is a natural cutoff that characterizes
the theory. The diagrammatic representation of (\ref{S4}) is
presented in Fig. 1.
\begin{figure}[hbt]
\hspace{-0.5cm}
\includegraphics[width=4.5cm,height=2cm]{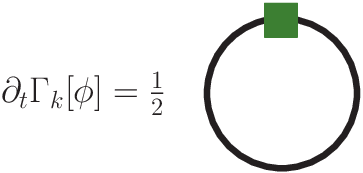}
\caption{Diagramatic representation of the exact RG
flow equation (\ref{S4}). The thick black line represents the full
connected two-point Green's function in the presence of the
additional IR cutoff ${\cal{R}}_{k}(q)$, i.e.
$G^{(2)}_{k}=[\Gamma_{k}^{(2)}+{\cal{R}}_{k}]^{-1}$.  The filled
green box represents the insertion of a factor
$\partial_{t}{\cal{R}}_{k}$.}
\end{figure}
\par
In this section, we will present the exact flow equation of the
effective average action of a cutoff $\lambda\varphi^{4}$ theory
with our cutoff procedure of (\ref{G4}). Our goal is to compare the
final form of the corresponding RG flow equation with (\ref{S4}).
Applications of the new RG flow equation and the consequences of the
new hidden gauge invariance, pointed out in Sec. III,  will be
presented elsewhere \cite{sadooghi}. To derive the above mentioned
flow equation, let us start by considering the action (\ref{G1}) in
the Euclidean space and to replace all fields $\tilde{\varphi}(q)$
in momentum space by $h_{k}(q)\tilde{\varphi}(q)$. In contrast to
the UV cutoff function $h_{\Lambda}(p)$, which satisfies
(\ref{G16}), the IR cutoff function $h_{k}(q)$ is considered to be a
sequence of analytic cutoff functions, converging to a sharp IR
cutoff $\Theta_{k,\Lambda}(q)$,\footnote{We will skip the subscripts
$\epsilon$ on $h_{\epsilon,k}(q)$ in the rest of this section.}
\begin{eqnarray}\label{S6-0}
h_{\epsilon,k}(q)\stackrel{\epsilon\to
0}{\longrightarrow}\Theta_{k,\Lambda}(q)=\left\{
\begin{array}{ccccrcl}
1&\qquad&\mbox{for}&\qquad&k^{2}&<& q^{2}<\Lambda^{2},\\
0&\qquad&\mbox{for}&\qquad&q^{2}&\leq&k^{2}.
\end{array}
\right.
\end{eqnarray}
Here, $\Lambda$ is an arbitrary UV cutoff, that cuts the UV modes
with $q>\Lambda$. Thus, in contrast to the Wetterich's method, where
the IR modes, with momenta smaller than the IR cutoff $k$, are
``screened in a mass-like fashion'', $m\sim k$ \cite{pawlowski}, as
described above, in our case, the IR modes are excluded from the
theory via an analytic IR cutoff function $h_{k}(q)$, satisfying
(\ref{S6-0}). Let us note that our cutoff procedure is similar to
the standard blocking procedure leading to the well-known
Wegner-Houghton RGE \cite{wegner1973} (see \cite{liao1992} for a
review). The difference is that instead of the standard sharp IR
cutoff, the IR cutoff function $h_{k}(q)$ has to be an analytic
function that converges to the sharp IR cutoff $\Theta_{k}(q)$. This
is indeed necessary, because otherwise a deformed theory with a
sharp cutoff defines a deformed product that does not satisfy the
desired associativity condition (\ref{A2}) [see our explanations in
the previous section]. With the replacement $\tilde{\varphi}(q)\to
h_{k}(q)\tilde{\varphi}(q)$, the modified cutoff action of a
$\lambda\varphi^{4}$ theory reads
\begin{eqnarray}\label{S6}
S_{k}[\varphi]&=&\frac{1}{2}\int\frac{d^{4}q_{1}}{(2\pi)^{4}}\frac{d^{4}q_{2}}{(2\pi)^{4}}{\cal{H}}_{k}^{(2)}(q_{1},q_{2})\tilde{\varphi}(q_{1})
\tilde{\varphi}(q_{2})\nonumber\\
&&\hspace{2cm}+ \int\frac{d^{4}q_{1}}{(2\pi)^{4}}\cdots
\frac{d^{4}q_{4}}{(2\pi)^{4}}{\cal{H}}_{k}^{(4)}(q_{1},\cdots,q_{4})\tilde{\varphi}(q_{1})\tilde{\varphi}(q_{2})\tilde{\varphi}(q_{3})\tilde{\varphi}(q_{4})
,
\end{eqnarray}
where the cutoff functions ${\cal{H}}_{k}^{(2)}$ and
${\cal{H}}_{k}^{(4)}$ are defined by
\begin{eqnarray}\label{S7}
{\cal{H}}_{k}^{(2)}(q_{1},q_{2})&\equiv&
h_{k}(q_{1})h_{k}(q_{2})(q_{1}^{2}+m^{2})\delta(q_{1}+q_{2}),\nonumber\\
{\cal{H}}_{k}^{(4)}(q_{1},q_{2},q_{3},q_{4})&\equiv&
\frac{\lambda}{4!}h_{k}(q_{1})h_{k}(q_{2})h_{k}(q_{3})
h_{k}(q_{4})\delta(q_{1}+q_{2}+q_{3}+q_{4}).
\end{eqnarray}
In App. B, we will follow the method described in
\cite{wetterich1993},\footnote{See also \cite{berges2000,
pawlowski}.} and will derive the corresponding exact RG flow
equation  to the effective average action arising from (\ref{S6}).
We will show that the RG flow equation of the effective average
action of the cutoff $\lambda \varphi^{4}$ theory is given by [see
also (\ref{appCC11})]
\begin{eqnarray}\label{S8}
\lefteqn{\frac{\partial\Gamma_{k}[\phi]}{\partial t}=\frac{1}{2}
\int \frac{d^{4}q_{1}}{(2\pi)^{4}}
\frac{d^{4}q_{2}}{(2\pi)^{4}}\frac{\partial
{\cal{H}}_{k}^{(2)}(q_{1},q_{2})}{\partial t}
[G_{k}^{(2)}(q_{1},q_{2})+\tilde{\phi}(q_{1})\tilde{\phi}(q_{2})]}\nonumber\\
&&+\int
\frac{d^{4}q_{1}}{(2\pi)^{4}}\cdots\frac{d^{4}q_{4}}{(2\pi)^{4}}
\frac{\partial {\cal{H}}_{k}^{(4)}(q_{1},\cdots,q_{4})}{\partial
t}\bigg[\tilde{\phi}(q_{1})\tilde{\phi}(q_{2})\tilde{\phi}(q_{3})\tilde{\phi}(q_{4})+
3G_{k}^{(2)}(q_{1},q_{2})G_{k}^{(2)}(q_{3},q_{4})\nonumber\\
&&\hspace{3cm}+6G_{k}^{(2)}(q_{1},q_{2})
\tilde{\phi}(q_{3})\tilde{\phi}(q_{4})+
4G_{k}^{(3)}(q_{1},q_{2},q_{3})\tilde{\phi}(q_{4})+G_{k}^{(4)}(q_{1},\cdots,q_{4})\bigg],
\end{eqnarray}
where the full two-point Green's function $G_{k}^{(2)}(p,q)$ are to
be replaced by $G_{k}^{(2)}(p,q)=G_{k}^{(2)}(q)\delta(p-q)$ with
$G^{(2)}_{k}(q)=[\Gamma_{k}^{(2)}(q)]^{-1}$. For the three- and
four-point Green's functions, $G_{k}^{(3)}$ and $G_{k}^{(4)}$ in the
remaining terms of (\ref{S8}), they will be replaced by
\begin{eqnarray}\label{S9}
G^{(3)}_{k}(q_{1},q_{2},q_{3})&=&-[\Gamma_{k}^{(2)}(q_{1})]^{-1}[\Gamma_{k}^{(2)}(q_{2})]^{-1}
[\Gamma_{k}^{(2)}(q_{3})]^{-1}\Gamma_{k}^{(3)}(q_{1},q_{2},q_{3}).\nonumber\\
G_{k}^{(4)}(q_{1},\cdots,q_{4})&=&-[\Gamma_{k}^{(2)}(q_{1})]^{-1}\cdots[\Gamma_{k}^{(2)}(q_{4})]^{-1}
\Gamma_{k}^{(4)}
(q_{1},q_{2},q_{3},q_{4})\nonumber\\
&&\hspace{-2cm}+3~[\Gamma_{k}^{(2)}(q_{1})]^{-1}\cdots[\Gamma_{k}^{(2)}(q_{4})]^{-1}\int\frac{d^{4}\ell}{(2\pi)^{4}}[\Gamma_{k}^{(2)}(\ell)]^{-1}
\Gamma_{k}^{(3)}(q_{1},q_{2},\ell)
\Gamma_{k}^{(3)}(\ell,q_{3},q_{4}).
\end{eqnarray}
The graphical representation of (\ref{S8}) is demonstrated in Fig.
2. At this stage a couple of remarks are in order. First, let us
notice that in the cutoff $\lambda\varphi^{4}$ theory, the relations
between the $n$-point Green's functions $G_{k}^{(n)}$ and $n$-point
vertex functions $\Gamma_{k}^{(n)}$ are not directly affected by the
cutoff function $h_{k}(q)$  [for a proof, see App. B]. This is in
contrast to, e.g. (\ref{S4a}), where the ordinary relation between
two-point Green's function and 1PI two-point vertex function is
modified with an additional term including the cutoff function
${\cal{R}}_{k}(q)$. As it turns out, this is because of the
multiplicative nature of $h_{k}(q)$, in contrast to the additive
nature of $\Delta S_{k}$ from (\ref{S1}), consisting of the IR
cutoff function ${\cal{R}}_{k}(q)$. The second point concerns the
appearance of new additional contributions in (\ref{S8}) compared to
(\ref{S4}). This is because in the cutoff $\lambda\varphi^{4}$
theory, $h_{k}(q)$ appears not only in the kinetic part of the
classical cut off action, as in the standard derivation of
(\ref{S4}), but also in interaction part of the classical action, as
all the new contributions are proportional to
$\partial_{t}{\cal{H}}_{k}^{(4)}$, with ${\cal{H}}_{k}^{(4)}$ from
(\ref{S7}), appearing in the interaction part of $S_{k}$ from
(\ref{S6}). It would be interesting to explore the practical
consequences of these new terms in RG flow equation (\ref{S8}). This
will be done elsewhere \cite{sadooghi}. Let us also note that the
procedure leading to (\ref{S8}) can be easily generalized to Abelian
and non-Abelian gauge theories. As we have shown in the previous
section, a new hidden gauge symmetry associated with the cutoff
procedure used in this paper exists, which guarantees the gauge
invariance along the flow equation. This is in contrast to the
situation of the standard Wetterich's exact RGE, where the manifest
gauge invariance is lost, because the regulator is not manifestly
gauge invariant \cite{pawlowski}.
\begin{figure}[t]
\includegraphics[width=16.5cm,height=5.5cm]{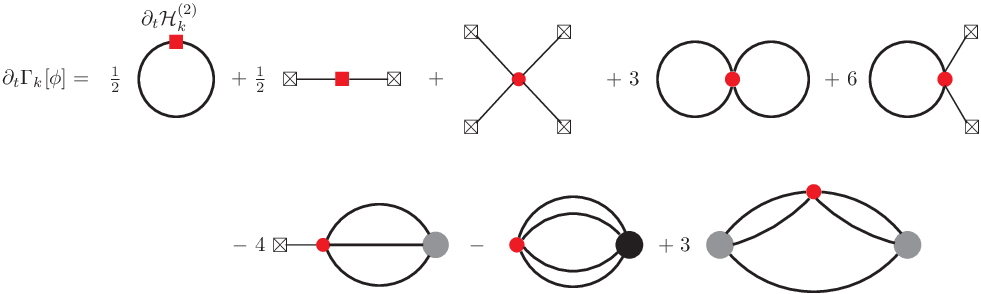}
\caption{Diagramatic representation of the RG flow
equation of the corresponding effective average action for cutoff
$\lambda\varphi^{4}$ theory (\ref{S8}). The thick black line
represents full two-point Green's functions
$G^{(2)}_{k}=[\Gamma_{k}^{(2)}]^{-1}$. The filled red boxes and the
small red circles represent the insertion of a factor
$\partial_{t}{\cal{H}}_{k}^{(2)}$ and
$\partial_{t}{\cal{H}}_{k}^{(4)}$, respectively. The thin lines
connected to $\boxtimes$ denote the background field
$\phi=\langle\varphi\rangle$. The big gray and black circles are 1PI
three- as well as four-point vertex functions, $\Gamma_{k}^{(3)}$ as
well as $\Gamma_{k}^{(4)}$, respectively. }
\end{figure}
\section{Conclusions}
\par\noindent
In the first part of this work we presented the procedure of
introducing a momentum cutoff in a field theory by directly cutting
off the momentum on each field via  a sequence of analytic UV cutoff
functions, $h_{\Lambda}(p)$, that converge appropriately to a sharp
UV cutoff $\Theta_{\Lambda}(p)$ from (\ref{G16}). For QED the
resulting exact renormalization group flow equation was shown to
respect a generalized ``cutoff gauge invariance'' which ensures
renormalizability, and unitarity without the need for an explicit
calculation using modified Ward-Takahashi identities.
\par
We need to emphasize  that although the gauge symmetry found is
motivated from a noncommutative geometric setup, this symmetry is an
inherent symmetry of ordinary QED cut off in momentum, in the spirit
of Polchinski's procedure. Our cutoff procedure may also be applied
to  non-Abelian gauge theories in a similar manner and the resultant
exact renormalization group flow be used for calculation of various
nonperturbative quantities in QCD \cite{pawlowski} and in gravity
\cite{RGgravity}.
\par
There is a subtle point that we would like to emphasize in
conclusion and that is the singular nature of our cutoff function
$h_{\Lambda}(p)$. As stated earlier, this function is quite general,
subject only to the restriction that it be equal to unity for
momenta smaller than $\Lambda$ and vanishing rapidly above it.
Strictly  speaking the cutoff function $h_{\Lambda}(p)$ is to be
chosen as a sequence of analytic functions, that converge to sharp
UV cutoff function $\Theta_{\Lambda}(p)$, defined in (\ref{G16}).
This is also recently indicated in \cite{lizzi2012}. Let us notice
again that $h_{\Lambda}(p)$ is to be smooth, as a sharp cutoff
function violates the associativity of our translationally invariant
star-algebra. Moreover, the symmetries and the correct
Ward-Takahashi's identities for each fixed $\Lambda$ are only
guaranteed when $h_{\Lambda}$ satisfies (\ref{G16}). In
\cite{lizzi2012}, it is also shown that the new deformed symmetry,
proposed in the present paper, results in a new Hopf algebra, which
is mathematically inequivalent with the undeformed one. This
guarantees that the new (deformed) cutoff gauge invariance is a new
genuine symmetry. This also resolves the puzzle that multiplication
of the fields by a function $h_{\Lambda}$, if it were smooth, would
simply be a field redefinition of the theory and therefore
physically trivial.
\par
In the second part of the paper, we used an analytic IR cutoff
function, $h_{k}(q)$, that converges, as its UV counterpart, to a
sharp IR cutoff function, $\Theta_{k}(q)$ defined in (\ref{S6-0}).
For this scale dependent cutoff function, $h_{k}$, the RG flow of
the effective average action is derived and is shown to be different
from the standard flow equation from \cite{wetterich1993,
morris1993, berges2000, pawlowski}. In \cite{sadooghi}, we will
generalize the method leading to the exact RG equation of cutoff
$\lambda\varphi^{4}$ theory to gauge theories and will explore the
practical consequences of the new generalized gauge symmetry
together with the effect of new terms appearing in the RG flow
equation (\ref{S8}) of the effective average action corresponding to
these cutoff gauge theories.
\section{Acknowledgments}
\par\noindent
The authors thank S. Rouhani, M. Alishahiha and A.~E. Mosaffa for
useful discussions.
\begin{appendix}
\section{The Proof of Ward identities} \setcounter{equation}{0}\noindent
To prove the Ward identities of cutoff QED, we start with the
generating functional for the full Green's functions
\begin{eqnarray}\label{B1}
Z[J_{\mu},\chi,\bar{\chi}]=\int
{\cal{D}}\psi~{\cal{D}}\bar{\psi}~{\cal{D}}A_{\mu}~e^{iS_{\mbox{\tiny{tot}}}},
\end{eqnarray}
with
$S_{\mbox{\tiny{tot}}}=S_{\mbox{\tiny{QED}}}+S_{\mbox{\tiny{source}}}$.
Here, $S_{\mbox{\tiny{QED}}}$ is given by (\ref{A9}), where the
ghost terms can be ignored, and $S_{\mbox{\tiny{source}}}$ by
\begin{eqnarray}\label{B2}
S_{\mbox{\tiny{source}}}=\int d^{4}x \left(\bar{\chi}\star
\psi+\bar{\psi}\star\chi+J_{\mu}\star A^{\mu}\right).
\end{eqnarray}
Varying $Z[J_{\mu},\chi,\bar{\chi}]$ in (\ref{B1}) with respect to
the star-gauge transformation (\ref{A11}) and replacing the
star-products with the expression on the r.h.s. of (\ref{A13}) to
introduce the cutoff function $h_{\Lambda}$, we arrive at
\begin{eqnarray}\label{B3}
&&\hspace{-1cm}\bigg\{-
 \frac{1}{\xi}p^{2}p^{\mu}h_{\Lambda}^2(p)\tilde{A}_{\mu}(p)+
p^{\mu}h_{\Lambda}^2(p)\tilde{J}_{\mu}(p)\nonumber\\
&&\hspace{-0.5cm}- eh_{\Lambda}(p)
\int\frac{d^{4}q}{(2\pi)^{4}}\bigg[h_{\Lambda}(p-q)\tilde{\bar\psi}(q-p)h_{\Lambda}(q)\tilde{\chi}(q)
-h_{\Lambda}(p-q)\tilde{\bar\chi}(q-p)h_{\Lambda}(q)\tilde{\psi}(q)\bigg]
\bigg\} Z=0.
\end{eqnarray}
Using the relations\footnote{We follow the notations in
\cite{pokorski}.}
\begin{eqnarray}\label{B4}
\tilde{\psi}(p)={h_{\Lambda}^{-2}(p)}\frac{\delta
Z}{i\delta\tilde{\bar{\chi}}(p)},\qquad \tilde{\bar{\psi}}(p)=
-h_{\Lambda}^{-2}(p)\frac{\delta Z}{i\delta\tilde{{\chi}}(p)},
\qquad \tilde{A}_{\mu}(p)=h_{\Lambda}^{-2}(p)\frac{\delta
Z}{i\delta\tilde{J}_{\mu}(-p)},
\end{eqnarray}
and replacing $Z[J_{\mu},\chi,\bar{\chi}]$ by $Z=e^{iW}$, with
$W[J_{\mu},\chi,\bar{\chi}]$ the generating function of connected
Green's functions, we arrive at
\begin{eqnarray}\label{B5}
&&\hspace{-1cm}-\frac{1}{\xi}p^{2}p_{\mu}\frac{\delta
W}{\delta\tilde{J}_{\mu}(-p)}+h^{2}_{\Lambda}(p)p_{\mu}\tilde{J}^{\mu}(p)
\nonumber\\
&&\hspace{-0.5cm}+eh_\Lambda(p)
\int\frac{d^{4}q}{(2\pi)^{4}}\bigg[h_{\Lambda}(q)h_{\Lambda}^{-1}(q-p)\frac{\delta
W}{\delta\tilde{\chi}(q-p)}\tilde{\chi}(q)+h_{\Lambda}^{-1}(q)h_{\Lambda}(q-p)\tilde{\bar{\chi}}(q-p)\frac{\delta
W}{\delta\tilde{\bar{\chi}}(q)} \bigg]=0.
\end{eqnarray}
To derive (\ref{B4}), $h_{\Lambda}(p)=h_{\Lambda}(-p)$ is used. To
arrive at the Ward identity in terms of
$\Gamma[A_{\mu},\psi,\bar{\psi}]$, the generating functional for 1PI
Green's function, we use the Legendre transformation (\ref{H10})
leading to
\begin{eqnarray}\label{B6}
&& \frac{\delta
W}{\delta\tilde{J}^{\mu}(-p)}=h_{\Lambda}^{2}(p)\tilde{A}_{\mu}(p),\qquad\qquad\hspace{0cm}
\frac{\delta
W}{\delta\tilde{\chi}(p)}=-h_{\Lambda}^{2}(p)\tilde{\bar\psi}(p),\qquad
~\frac{\delta
W}{\delta\tilde{\bar\chi}(p)}=h_{\Lambda}^{2}(p)\tilde{\psi}(p),\nonumber\\
&&\ \ \tilde{J}_{\mu}(p)=-h_{\Lambda}^{-2}(p)\frac{\delta
\Gamma}{\delta \tilde{A}^{\mu}(-p)},\qquad\ \ \
\tilde{\chi}(p)=-h_{\Lambda}^{-2}(p)\frac{\delta \Gamma}{\delta
\tilde{\bar\psi}(p)},\qquad
\tilde{\bar\chi}(p)=h_{\Lambda}^{-2}(p)\frac{\delta \Gamma}{\delta
\tilde{\psi}(p)}.
\end{eqnarray}
Plugging these relations in (\ref{B5}), we arrive at
\begin{eqnarray}\label{B7}
&&\hspace{-1cm}-\frac{1}{\xi}p^{2}p_{\mu}h^{2}_{\Lambda}(p)A^{\mu}(p)-
p_{\mu}\frac{\delta\Gamma}{\delta\tilde{A}_{\mu}(-p)}
\nonumber\\
&&\hspace{-0.3cm}+eh_\Lambda(p)
\int\frac{d^{4}q}{(2\pi)^{4}}\bigg[h^{-1}_{\Lambda}(q)h_{\Lambda}(q-p)\tilde{\bar{\psi}}(q-p)
\frac{\delta\Gamma}{\delta\tilde{\bar{\psi}}(q)}
+h_{\Lambda}(q)h^{-1}_{\Lambda}(q-p)\frac{\delta\Gamma}{\delta\tilde{\psi}(q-p)}\tilde{\psi}(q)
\bigg]=0.
\end{eqnarray}
As a first example on the application of (\ref{B5}), let us
differentiate it with respect to $\tilde{J}_{\nu}(p')$ and set
eventually $\tilde{J}_{\mu}=\tilde{\bar{\chi}}=\tilde{\chi}=0$. We
arrive at the Ward identity for the full photon propagator of the
cutoff QED in momentum space, $\tilde{D}^{\mu\nu}_{\Lambda}$,
\begin{eqnarray}\label{B8}
\frac{i}{\xi}p^{2}p_{\mu}h_{\Lambda}^{2}(p)\tilde{D}_{\Lambda}^{\mu\nu}(p)=p^{\nu}.
\end{eqnarray}
Here, we have used (\ref{B6}) to get first
\begin{eqnarray}\label{B9}
\frac{\delta
W}{\delta\tilde{J}_{\mu}(-p)\delta\tilde{J}_{\nu}(p')}\bigg|_{\tilde{J}_{\mu}=\tilde{\bar{\chi}}=\tilde{\chi}=0}=
h_{\Lambda}^{4}(p)\langle
\tilde{A}^{\mu}(p)\tilde{A}^{\nu}(-p')\rangle,
\end{eqnarray}
and defined the full cutoff dependent photon propagator
$\tilde{D}_{\Lambda}^{\mu\nu}(p)$ by
\begin{eqnarray}\label{B10}
i\tilde{D}^{\mu\nu}_{\Lambda}(p)\delta(p-p')\equiv
\langle\tilde{A}^{\mu}(p)\tilde{A}^{\nu}(-p')\rangle.
\end{eqnarray}
Equation (\ref{B8}) is in particular satisfied by tree level photon
propagator of cutoff QED \cite{ardalan2010}
\begin{eqnarray}\label{B11}
\tilde{D}^{\mu\nu}_{\Lambda}(p)\bigg|_{\mbox{\tiny{tree-level}}}=-\frac{ih_{\Lambda}^{-2}(p)}{p^{2}}
\left(g^{\mu\nu}-(1-\xi)\frac{p^{\mu}p^{\nu}}{p^{2}}\right).
\end{eqnarray}
As a second example, let us differentiate (\ref{B7}) with respect to
$\tilde{\psi}(-\ell)$ and $\tilde{\bar{\psi}}(k)$ and set eventually
$\tilde{A}_{\mu}=\tilde{\psi}=\tilde{\bar{\psi}}=0$. Using
(\ref{B6}), we arrive at
\begin{eqnarray}\label{B12}
\frac{1}{h_{\Lambda}(p)h_{\Lambda}(\ell+p)h_{\Lambda}(\ell)}p_{\mu}\tilde{\Gamma}^{\mu}_{\Lambda}(-\ell-p,-\ell;-p)&=&e\bigg[h^{-2}_{\Lambda}(\ell+p)
\tilde{S}_{\Lambda}^{-1}(\ell+p)-h^{-2}_{\Lambda}(\ell)\tilde{S}_{\Lambda}^{-1}(\ell)\bigg],
\end{eqnarray}
where the 1PI three point vertex
 function
\begin{eqnarray}\label{B13}
\tilde{\Gamma}^{\mu}_{\Lambda}(k,-\ell;-p)\delta(k+p+\ell)\equiv\frac{\delta^{3}\Gamma}{\delta\tilde{\bar{\psi}}(k)\delta\tilde{\psi}(-\ell)\delta
\tilde{A}_{\mu}(-p)}\bigg|_{\tilde{A}_{\mu}=\tilde{\psi}=\tilde{\bar{\psi}}=0},
\end{eqnarray}
as well the fermionic 1PI two-point function at finite cutoff
$\Lambda$
\begin{eqnarray}\label{B14}
\tilde{S}_{\Lambda}^{-1}(\ell)\delta(k+\ell)\equiv
\frac{\delta^{2}\Gamma}{\delta\tilde{\bar{\psi}}(k)\delta\tilde{\psi}(-\ell)}\bigg|_{\tilde{\psi}=\tilde{\bar{\psi}}=0},
\end{eqnarray}
are introduced. Taking the limit $p\to 0$ in (\ref{B12}) and using
$h_{\Lambda}(0)=1$, we arrive at the standard relation
\begin{eqnarray}\label{B15}
\tilde{\Gamma}^{\mu}_{\Lambda}(-\ell,-\ell;0)=e\frac{\partial
\tilde{S}_{\Lambda}^{-1}(\ell)}{\partial\ell_{\mu}}.
\end{eqnarray}
Assuming that $h_{\Lambda}(p)$ is a nearly constant function for
$|p|<\Lambda$, and using\footnote{Relations (\ref{B16}) are shown to
be valid at one-loop level (see \cite{ardalan2010} for more
details).}
\begin{eqnarray}\label{B16}
\tilde{S}_{\Lambda}^{-1}(\ell)&=&h_{\Lambda}^{2}(\ell)\tilde{S}_{\infty}^{-1}(\ell),\nonumber\\
\tilde{\Gamma}^{\mu}_{\Lambda}(k,\ell;p)&=&h_{\Lambda}(k)h_{\Lambda}(\ell)h_{\Lambda}(p)\tilde{\Gamma}^{\mu}_{\infty}(k,\ell;p),
\end{eqnarray}
we get
\begin{eqnarray}\label{B17}
\tilde{\Gamma}^{\mu}_{\infty}(-\ell,-\ell;0)=e\frac{\partial
\tilde{S}_{\infty}^{-1}(\ell)}{\partial\ell_{\mu}},
\end{eqnarray}
where $\tilde{S}_{\infty}$ and $\tilde{\Gamma}_{\infty}^{\mu}$ are
the 1PI two- and three-point vertex functions of QED in $\Lambda\to
\infty$ limit. Assuming at this stage that (\ref{B17}) is also valid
for renormalized Green's functions
$[\tilde{\Gamma}^{\mu}_{\infty}]_{r}=Z_{2}^{-1}Z_{3}^{-1/2}\tilde{\Gamma}^{\mu}_{\infty}$
and
$[\tilde{S}_{\infty}^{-1}]_{r}=Z_{2}^{-1}\tilde{S}_{\infty}^{-1}$,
as well as for renormalized coupling $e_{r}\equiv
Z_{1}Z_{2}^{-1}Z_{3}^{-1/2}e$, with $Z_{1}, Z_{2}$ and $Z_{3}$ the
renormalization constants corresponding to the vertex function,
fermion and photon propagators, respectively, we arrive at
$Z_{1}=Z_{2}$.
\section{Exact RG flow equation of cutoff $\lambda\varphi^{4}$ theory}\label{appC}
\setcounter{equation}{0}\noindent
Let us start by considering the bare action of $\lambda\varphi^{4}$
theory in Euclidean space (\ref{G1}). As we have explained in Sec.
\ref{sec3}, each field $\tilde{\varphi}(q)$ shall be replaced by
$h_{k}(q)\varphi(q)$, where $k$ is the renormalization scale. The
modified classical action is then given by (\ref{S6}) with the
cutoff functions given in (\ref{S7}). The corresponding generating
functional of this cutoff theory then reads
\begin{eqnarray}\label{appCC1}
Z_{k}[J]=\int {\cal{D}}\varphi~\exp\left(-S_{k}[\varphi]+\int
\frac{d^{4}q}{(2\pi)^{4}}h_{k}^{2}(q)\tilde{J}(-q)\tilde{\varphi}(q)\right),
\end{eqnarray}
where $h_{k}(-q)=h_{k}(q)$ is assumed. The Legendre transformation
between $W_{k}[J]\equiv \ln Z_{k}[J]$, the generating functional of
the connected Green's function, and the 1PI effective average
action, $\Gamma_{k}[\phi]$ is given by
\begin{eqnarray}\label{appCC2}
\Gamma_{k}[\phi]=-W_{k}[J]+\int
\frac{d^{4}q}{(2\pi)^{4}}h_{k}^{2}(q)\tilde{J}(-q)\tilde{\phi}(q),
\end{eqnarray}
where $\phi\equiv \langle\varphi\rangle$, and
\begin{eqnarray}\label{appCC3}
\frac{\delta W_{k}[J]}{\delta
\tilde{J}(-q)}=h_{k}^{2}(q)\tilde{\phi}(q).
\end{eqnarray}
It is the purpose of this appendix to derive the scale dependence of
$\Gamma_{k}[\phi]$. To do this, we will follow the method described
in \cite{wetterich1993, berges2000,pawlowski}. First, we
differentiate $\Gamma_{k}[\phi]$ from (\ref{appCC2}) with respect to
$k$ and arrive at
\begin{eqnarray}\label{appCC4}
\frac{\partial\Gamma_{k}[\phi]}{\partial k}=-\frac{\partial
W_{k}[J]}{\partial
k}+2\int\frac{d^{4}q}{(2\pi)^{4}}h_{k}(q)\frac{\partial
h_{k}(q)}{\partial
k}\tilde{J}(-q)\tilde{\phi}(q)=\bigg\langle\frac{\partial
S_{k}[\varphi]}{\partial k}\bigg\rangle,
\end{eqnarray}
where (\ref{appCC3}) and the standard notation
\begin{eqnarray}\label{appCC5}
\langle{\cal{O}}[\varphi]\rangle=Z_{k}^{-1}[J]\int
{\cal{D}}\varphi~{\cal{O}}[\varphi]\ \exp\left(-S_{k}[\varphi]+\int
\frac{d^{4}q}{(2\pi)^{4}}h_{k}^{2}(q)\tilde{J}(-q)\tilde{\varphi}(q)\right),
\end{eqnarray}
are used. Plugging now (\ref{S6}) in (\ref{appCC4}), we get
\begin{eqnarray}\label{appCC6}
\frac{\partial\Gamma_{k}[\phi]}{\partial
k}=\bigg\langle\frac{\partial S_{k}[\varphi]}{\partial
k}\bigg\rangle
&=&\frac{1}{2}\int\frac{d^{4}q_{1}}{(2\pi)^{4}}\frac{d^{4}q_{2}}{(2\pi)^{4}}\frac{\partial
{\cal{H}}_{k}^{(2)}(q_{1},q_{2})}{\partial
k}\langle\tilde{\varphi}(q_{1})\tilde{\varphi}(q_{2})\rangle\nonumber\\
&&+ \int\frac{d^{4}q_{1}}{(2\pi)^{4}}\cdots
\frac{d^{4}q_{4}}{(2\pi)^{4}}\frac{\partial
{\cal{H}}_{k}^{(4)}(q_{1},\cdots,q_{4})}{\partial
k}\langle\tilde{\varphi}(q_{1})\tilde{\varphi}(q_{2})\tilde{\varphi}(q_{3})\tilde{\varphi}(q_{4})\rangle
.
\end{eqnarray}
At this stage we shall replace the two- and four-point Green's
functions appearing on the r.h.s. of (\ref{appCC6}) by a combination
of connected and disconnected Green's functions. To do this, let us
vary $W_{k}[J]=\ln Z_{k}[J]$ two times with respect to $\tilde{J}$
to get
\begin{eqnarray}\label{appCC7}
\frac{\delta^{2}W_{k}[J]}{\delta\tilde{J}(-q_{1})\delta\tilde{J}(-q_{2})}&=&-\frac{1}{Z_{k}^{2}[J]}\frac{\delta
Z_{k}[J]}{\delta\tilde{{J}}(-q_{1})}\frac{\delta
Z_{k}[J]}{\delta\tilde{J}(-q_{2})}+\frac{1}{Z_{k}[J]}\frac{\delta^{2}
Z_{k}[J]}{\delta\tilde{J}(-q_{1})\delta\tilde{J}(-q_{2})}\nonumber\\
&=&h_{k}^{2}(q_{1})h_{k}^{2}(q_{2})\left(-\tilde{\phi}(q_{1})\tilde{\phi}(q_{2})+\langle\tilde{\varphi}(q_{1})
\tilde{\varphi}(q_{2})\rangle\right).
\end{eqnarray}
Here, the definition of $Z_{k}[J]$ from (\ref{appCC1}) is used.
Defining then the connected $n$-point Green's functions as
\begin{eqnarray}\label{appCC8}
\frac{\delta^{n}W_{k}[J]}{\delta\tilde{J}(-q_{1})\cdots\tilde{J}(-q_{n})}\equiv
h_{k}^{2}(q_{1})\cdots h_{k}^{2}(q_{n})G_{k}^{(n)}(q_{1},\cdots,
q_{n}),
\end{eqnarray}
and plugging the corresponding relation for $n=2$ on the l.h.s. of
(\ref{appCC7}), we get the standard relation
\begin{eqnarray}\label{appCC9}
\langle\tilde{\varphi}(q_{1})
\tilde{\varphi}(q_{2})\rangle=G_{k}^{(2)}(q_{1},q_{2})+\tilde{\phi}(q_{1})\tilde{\phi}(q_{2}).
\end{eqnarray}
Similar relation exists also between the connected four-point
Green's function $G_{k}^{(4)}(q_{1},\cdots,q_{4})$ and
$\langle\tilde{\varphi}(q_{1})\tilde{\varphi}(q_{2})\tilde{\varphi}(q_{3})\tilde{\varphi}(q_{4})\rangle
$ appearing on the r.h.s. of (\ref{appCC6}). It is given by
\begin{eqnarray}\label{appCC10}
\lefteqn{\langle\tilde{\varphi}(q_{1})\tilde{\varphi}(q_{2})\tilde{\varphi}(q_{3})\tilde{\varphi}(q_{4})\rangle=
G_{k}^{(4)}(q_{1},\cdots,q_{4})
}\nonumber\\
&&+G_{k}^{(3)}(q_{1},q_{2},q_{3})\tilde{\phi}(q_{4})+G_{k}^{(3)}(q_{1},q_{2},q_{4})\tilde{\phi}(q_{3})+
G_{k}^{(3)}(q_{1},q_{4},q_{3})\tilde{\phi}(q_{2})+G_{k}^{(3)}(q_{4},q_{2},q_{3})\tilde{\phi}(q_{1})\nonumber\\
&&+G_{k}^{2}(q_{1},q_{2})\tilde{\phi}(q_{3})\tilde{\phi}(q_{4})
+G_{k}^{2}(q_{1},q_{3})\tilde{\phi}(q_{2})\tilde{\phi}(q_{4})
+G_{k}^{2}(q_{1},q_{4})\tilde{\phi}(q_{2})\tilde{\phi}(q_{3})
+G_{k}^{2}(q_{2},q_{3})\tilde{\phi}(q_{1})\tilde{\phi}(q_{4})\nonumber\\
&& +G_{k}^{2}(q_{2},q_{4})\tilde{\phi}(q_{1})\tilde{\phi}(q_{3})
+G_{k}^{2}(q_{3},q_{4})\tilde{\phi}(q_{1})\tilde{\phi}(q_{2})
\nonumber\\
 &&+G_{k}^{(2)}(q_{1},q_{2})
G_{k}^{(2)}(q_{3},q_{4})+G_{k}^{(2)}(q_{1},q_{3})
G_{k}^{(2)}(q_{2},q_{4})+G_{k}^{(2)}(q_{1},q_{4})
G_{k}^{(2)}(q_{2},q_{3})\nonumber\\
&&+\tilde{\phi}(q_{1})\tilde{\phi}(q_{2})\tilde{\phi}(q_{3})\tilde{\phi}(q_{4}).
\end{eqnarray}
Plugging (\ref{appCC9}) and (\ref{appCC10}) in (\ref{appCC6}) and
using the symmetry of ${\cal{H}}_{k}^{(2)}(q_{1},q_{2})$ and
${\cal{H}}_{k}^{(4)}(q_{1},\cdots,q_{4})$ under permutation of
$q_{i},i=1,\cdots,4$, we arrive at the flow equation of
$\Gamma_{k}[\phi]$ in terms of the connected $n=1,\cdots,4$-point
Green's functions
\begin{eqnarray}\label{appCC11}
\lefteqn{\frac{\partial\Gamma_{k}[\phi]}{\partial t}=\frac{1}{2}
\int \frac{d^{4}q_{1}}{(2\pi)^{4}}
\frac{d^{4}q_{2}}{(2\pi)^{4}}\frac{\partial
{\cal{H}}_{k}^{(2)}(q_{1},q_{2})}{\partial t}
[G_{k}^{(2)}(q_{1},q_{2})+\tilde{\phi}(q_{1})\tilde{\phi}(q_{2})]}\nonumber\\
&&+\int
\frac{d^{4}q_{1}}{(2\pi)^{4}}\cdots\frac{d^{4}q_{4}}{(2\pi)^{4}}
\frac{\partial {\cal{H}}_{k}^{(4)}(q_{1},\cdots,q_{4})}{\partial
t}\bigg[\tilde{\phi}(q_{1})\tilde{\phi}(q_{2})\tilde{\phi}(q_{3})\tilde{\phi}(q_{4})+
3G_{k}^{(2)}(q_{1},q_{2})G_{k}^{(2)}(q_{3},q_{4})\nonumber\\
&&\hspace{3cm}+6G_{k}^{(2)}(q_{1},q_{2})
\tilde{\phi}(q_{3})\tilde{\phi}(q_{4})+
4G_{k}^{(3)}(q_{1},q_{2},q_{3})\tilde{\phi}(q_{4})+G_{k}^{(4)}(q_{1},\cdots,q_{4})\bigg].
\end{eqnarray}
This flow equation is to be compared with the standard flow equation
(\ref{S4}), where only a term similar to the first term on the
r.h.s. of (\ref{appCC11}) appears. The appearance of additional
terms in (\ref{appCC11}), including the contributions of
$G_{k}^{(n)}(q_{1},\cdots,q_{n}), n=1,\cdots,4$, is, in particular,
a consequence of the replacement of $\tilde{\varphi}(q)$ by
$h_{k}(q)\tilde{\varphi}(q)$ in the interaction term of the original
classical action, in contrast to the standard procedure
\cite{wetterich1993, morris1993, berges2000, pawlowski}.
\par
In a last step, we shall use the relations between the connected
$n$-point Green's function, $G_{k}^{(n)}(q_{1},\cdots,q_{n})$,
defined in (\ref{appCC8}) and the 1PI $n$-point vertex functions,
$\Gamma_{k}^{(n)}(q_{1},\cdots,q_{n})$, defined by
\begin{eqnarray}\label{appCC12}
\Gamma_{k}^{(n)}(q_{1},\cdots,q_{n})\equiv\frac{\delta^{2}\Gamma_{k}[\phi]}
{\delta\tilde{\phi}(q_{1})\cdots\delta\tilde{\phi}(q_{n})},
\end{eqnarray}
to replace $G_{k}^{(n)}(q_{1},\cdots,q_{n}), n=1,\cdots,4$ in
(\ref{appCC11}) by the corresponding expressions in terms of
$\Gamma_{k}^{(n)}(q_{1},\cdots,q_{n})$. To do this, let us first
consider the relation
\begin{eqnarray}\label{appCC13}
\int \frac{d^{4}\ell_{1}}{(2\pi)^{4}}\frac{\delta
\tilde{J}(-q_{1})}{\delta\tilde{\phi}(\ell_{1})}\frac{\delta\tilde{\phi}(\ell_{1})}{\delta\tilde{J}(-q_{2})}=
\delta(q_{1}-q_{2}).
\end{eqnarray}
It can easily be shown that
\begin{eqnarray}\label{appCC14}
\frac{\delta
\tilde{J}(-q_{1})}{\delta\tilde{\phi}(\ell_{1})}=h_{k}^{-2}(q_{1})\Gamma_{k}^{(2)}(q_{1},\ell_{1}),
\qquad\mbox{and}\qquad
\frac{\delta\tilde{\phi}(\ell_{1})}{\delta\tilde{J}(-q_{2})}=h_{k}^{2}(q_{2})G^{(2)}_{k}(\ell_{1},q_{2}).
\end{eqnarray}
Plugging (\ref{appCC14}) in (\ref{appCC13}), and using
$G_{k}^{(2)}(\ell_{1},q_{2})=G_{k}^{(2)}(q_{2})\delta(\ell_{1}-q_{2})$
as well as $\Gamma_{k}^{(2)}(q_{1},\ell_{1})=
\Gamma_{k}^{(2)}(q_{1})\delta(q_{1}-\ell_{1})$, we arrive, after
integrating over $\ell_{1}$, at
\begin{eqnarray}\label{appCC15}
G_{k}^{(2)}(q)\Gamma_{k}^{(2)}(q)=1.
\end{eqnarray}
This is the standard relation between $G_{k}^{(2)}(q)$ and
$\Gamma_{k}^{(2)}(q)$. In contrast to (\ref{S4a}), the cutoff
function $h_{k}(q)$ does not appear in (\ref{appCC15}). This is
because of the multiplicative nature of the cutoff function
$h_{k}(q)$. Similar relation can also be derived between
$G_{k}^{(3)}$ and $\Gamma_{k}^{(3)}$. It is simplify given by
\begin{eqnarray}\label{appCC16}
G_{k}^{(3)}(q_{1},q_{2},q_{3})=-[\Gamma_{k}^{(2)}(q_{1})]^{-1}[\Gamma_{k}^{(2)}(q_{2})]^{-1}
[\Gamma_{k}^{(2)}(q_{3})]^{-1}\Gamma_{k}^{(3)}(q_{1},q_{2},q_{3}),
\end{eqnarray}
which is derived by differentiating (\ref{appCC13}) with respect to
$\tilde{J}(-q_{3})$, and plugging (\ref{appCC14}) as well as
\begin{eqnarray}\label{appCC17}
\frac{\delta^{2}
\tilde{J}(-q_{1})}{\delta\tilde{\phi}(\ell_{1})\delta\tilde{J}(-q_{3})}&=&h_{k}^{-2}(q_{1})h_{k}^{2}(q_{3})\int
\frac{d^{4}\ell_{2}}{(2\pi)^{4}}G_{k}^{(2)}(\ell_{2},q_{3})
\Gamma_{k}^{(3)}(q_{1},\ell_{1},\ell_{2}),\nonumber\\
\frac{\delta^{2}\tilde{\phi}(\ell_{1})}{\delta\tilde{J}(-q_{2})\delta\tilde{J}(-q_{3})}&=&
h_{k}^{2}(q_{2})h_{k}^{2}(q_{3})G_{k}^{(3)}(\ell_{1},q_{2},q_{3}).
\end{eqnarray}
in the resulting expression. Similarly, to determine the relation
between $G_{k}^{(4)}$ and $\Gamma_{k}^{(4)}$, we differentiate
(\ref{appCC13}) with respect to $\tilde{J}(-q_{3})$ and
$\tilde{J}(-q_{4})$. Plugging (\ref{appCC14}), (\ref{appCC17}) and
\begin{eqnarray}\label{appCC18}
\frac{\delta^{3}
\tilde{J}(-q_{1})}{\delta\tilde{\phi}(\ell_{1})\delta\tilde{J}(-q_{3})\delta\tilde{J}(-q_{4})}
&=&h_{k}^{-2}(q_{1})h_{k}^{2}(q_{3})h_{k}^{2}(q_{4})\bigg\{\int\frac{d^{4}\ell_{2}}{(2\pi)^{4}}
G_{k}^{(3)}(\ell_{2},q_{3},q_{4})
\Gamma_{k}^{(3)}(q_{1},\ell_{1},\ell_{2})\nonumber\\
&&+\int
\frac{d^{4}\ell_{2}}{(2\pi)^{4}}\frac{d^{4}\ell_{3}}{(2\pi)^{4}}G_{k}^{(2)}(\ell_{2},q_{3})G_{k}^{(2)}(\ell_{3},q_{4})
\Gamma_{k}^{(4)}(q_{1},\ell_{1},\ell_{2},\ell_{3})\bigg\},\nonumber\\
\frac{\delta^{3}\tilde{\phi}(\ell_{1})}{\delta\tilde{J}(-q_{2})
\delta\tilde{J}(-q_{3})\delta\tilde{J}(-q_{4})}&=&h_{k}^{2}(q_{2})h_{k}^{2}(q_{3})h_{k}^{2}(q_{4})
G_{k}^{(4)}(\ell_{1},q_{2},q_{3},q_{4}),
\end{eqnarray}
in the resulting expression, we arrive after some algebra at
\begin{eqnarray}\label{appCC19}
\lefteqn{G_{k}^{(4)}(q_{1},\cdots,q_{4})=-[\Gamma_{k}^{(2)}(q_{1})]^{-1}\cdots[\Gamma_{k}^{(2)}(q_{4})]^{-1}
\bigg\{\Gamma_{k}^{(4)} (q_{1},q_{2},q_{3},q_{4})
}\nonumber\\
&&-\int\frac{d^{4}\ell}{(2\pi)^{4}}[\Gamma_{k}^{(2)}(\ell)]^{-1}\bigg[\Gamma_{k}^{(3)}(q_{1},q_{2},\ell)
\Gamma_{k}^{(3)}(\ell,q_{3},q_{4})+\Gamma_{k}^{(3)}(q_{1},q_{3},\ell)\Gamma_{k}^{(3)}(\ell,q_{2},q_{4})\nonumber\\
&&+
\Gamma_{k}^{(3)}(q_{1},q_{4},\ell)\Gamma_{k}^{(3)}(\ell,q_{2},q_{3})\bigg]
\bigg\},
\end{eqnarray}
where (\ref{appCC15}) and (\ref{appCC16}) are also used. Plugging
(\ref{appCC15}), (\ref{appCC16}) and (\ref{appCC19}) in the flow
equation (\ref{appCC11}) and using the symmetry of the cutoff
function ${\cal{H}}_{k}^{(4)}(q_{1},\cdots, q_{4})$ under
permutation of $q_{i},i=1,\cdots,4$, we arrive at (\ref{S8}).
\end{appendix}

\end{document}